\documentclass[prl,twocolumn,superscriptaddress,showpacs,preprintnumbers,floatfix,amsmath,amssymb,nofootinbib,tightenlines]{revtex4}
\usepackage{graphicx}

\newcommand{\beqn}{\begin{eqnarray}}
\newcommand{\eeqn}{\end{eqnarray}}

\begin{document}


\title{The Single Flavor Color Superconductivity in a Magnetic Field}

\author{Bo Feng}
\affiliation{Institute of Particle Physics, Huazhong Normal
University, Wuhan 430079, China}
\author{De-fu Hou}
\affiliation{Institute of Particle Physics, Huazhong Normal
University, Wuhan 430079, China}
\author{Hai-cang Ren}\affiliation{Physics Department, The Rockefeller University,
1230 York Avenue, New York, NY 10021-6399}\affiliation{Institute
of Particle Physics, Huazhong Normal University, Wuhan 430079,
China}
\author{Ping-ping Wu}\affiliation{Institute
of Particle Physics, Huazhong Normal University, Wuhan 430079,
China}

\begin{abstract}
We investigate the single flavor color superconductivity in a
magnetic field. Because of the absence of the electromagnetic
Meissner effect, forming a nonspherical CSC phase, polar, A or
planar, does not cost energy of excluding magnetic flux. We found
that these nonspherical phases do occupy a significant portion of
the phase diagram with respect to magnetic field and temperature and
may be implemented under the typical quark density and the magnetic
field inside a neutron star.

\end{abstract}


\maketitle

A cold quark matter will become a color superconductor at
sufficiently high baryon density\cite{MAKT}. In the core region of
a compact star, the baryon density is expected to be several times
higher than that of a normal nuclear matter. The quarks may be
released from hadrons and form a quark matter of $\mu\sim
400-500$MeV, providing an opportunity to the color superconductivity(CSC).



While the pairing force is maximized in the s-wave channel, the
antisymmetry of the wave function requires the Cooper pairing
between different quark flavors. But the mass of strange
quarks and the charge neutrality induces a substantial Fermi
momentum mismatch among different flavors and thereby reduces the
phase space available for pairing. A number of exotic 2 flavor or
3 flavor CSC phases have been proposed without reaching a concensus
solution. The single flavor CSC (pairing within each flavor)
becomes a potential candidate even at disadvantage of a reduced pairing
strength. The dominant angular momentum of the single flavor Cooper pair is
one. Like the superfluidity of $^3$He, there are a number of different pairing
states and we shall focus in this letter the four of them: the
spherical color-spin locked (CSL)\cite{T,SWR} state and
nonspherical polar, A and planar ones. Without a magnetic field,
the CSL pairing is energetically most favored, even when the
angular momentum mixing effect is taken into account\cite{BDH}.

The energy balance among different single flavor CSC phases will be
offset in a magnetic field, which is present in a compact star and
could exceed $10^{15}$G in magnitude. Only the CSL phase shield the
magnetic field \cite{SWR}. The electromagnetic Meissner effect is
absent for nonspherical states (polar, A or planar). Cooling a
normal quark matter to the CSL will costs an extra amount work to
expel out the magnetic flux. Being free from such a penalty,
nonspherical phases may show up at a  sufficiently high  magnetic
field. Obtaining the phase diagram of a single flavor CSC with
respect to temperature and magnetic field is the main scope of the
present letter.

\begin{table*}
\caption{\label{tab:table3}This table shows possible phases under
a magnetic field for both two flavors and three
flavors cases with each flavor forming spin-one CSC or remaining
normal state. The scale of critical magnetic field and the
critical temperature have also been included.}
\begin{ruledtabular}
\begin{tabular}{ccccccc}
&I&II&III&IV&$H_0(10^{14
}G)$&$T_C(10^{-1}MeV)$\\
\hline
 2 flavor&$\rm CSL_u, CSL_d$&$\rm (polar)_u, (planar)_d$&$\rm (normal)_u, (polar)_d$&$\rm (normal)_u, (normal)_d$&5.44&1.35 \\
 3 flavor&$\rm CSL_u, CSL_{d,s}$&$\rm (polar)_u, (planar)_{d,s}$&$\rm (normal)_u, (polar)_{d,s}$&$\rm (normal)_u, (normal)_{d,s}$&1.97&0.49\\
\end{tabular}
\end{ruledtabular}
\end{table*}

The structure of the Meissner effect in a single flavor pairing is
determined by the pattern of its symmetry breaking\cite{SWR}. The
condensate of a diquark operator  takes the form
\begin{equation}
\Phi=<\bar\psi_C\Gamma^c\lambda^c\psi> \label{diquark}
\end{equation}
where $\psi$ is the quark field, $\psi_C=\gamma_2\psi^*$ is its
charge conjugate, $\lambda^c$ with $c=2,5,7$ is an antisymmetric
Gell-Mann matrices and $\Gamma^c$ is a $4\times 4$ spinor matrix.
We may choose $\Gamma^5=\Gamma^7=0$ for the polar and A phases,
$\Gamma^2=0$ for the planar phase but none of $\Gamma^c$'s
vanishes for CSL phase. The condensate of CSL breaks the gauge
symmetry $SU(3)_c\times U(1)_{\rm em.}$ completely. A nonspherical
condensate, however, breaks the gauge symmetry partially and the
Meissner effect is incomplete. Among the residual gauge group,
there exists a U(1) transformation, $\psi\to
e^{-\frac{i}{2}\lambda_8\theta-iq\phi}\psi$ with $q$ the electric
charge of $\psi$, $\theta=-2\sqrt{3}q\phi$ for the polar and A
phases and $\theta=4\sqrt{3}q\phi$ for the planar phase. The
corresponding gauge field, ${\cal A}_\mu$ is identified with the
electromagnetic field in the condensate. It is related to the
electromagnetic field $A$ and the 8-th component of the color
field $A^8$ in the normal phase through a rotation
\begin{eqnarray}
{\cal A}_\mu &=& A_\mu\cos\gamma-A_\mu^8\sin\gamma\nonumber\\
{\cal V}_\mu &=& A_\mu\sin\gamma+A_\mu^8\cos\gamma\label{transform}
\end{eqnarray}
where $\tan\gamma=-2\sqrt{3}q(e/g)$ for polar and A, and
$\tan\gamma=4\sqrt{3}q(e/g)$ for planar with $g$ the QCD running
coupling constant. The 2nd component of (\ref{transform}) ${\cal
V}=0$ because of the Meissner effect and thereby imposes a
constraint inside a nonspherical CSC, $A_\mu^8=-A_\mu\tan\gamma$,
which implies the relation\cite{relation}
\begin{equation}
{\bf B}^8=-{\bf B}\tan\gamma
\label{constraint}
\end{equation}
between the color and the ordinary magnetic fields. Expressing the gauge coupling
\begin{equation}
\bar\psi\gamma_\mu(eqA_\mu+\frac{1}{2}A_\mu^8\lambda_8)\psi
\end{equation}
in terms of ${\cal A}_\mu$ and its orthogonal partner ${\cal
V}_\mu$, we extract the electric charges with respect to ${\cal
A}$ in color space,
\begin{equation}
Q=\left\{\begin{array}{ll}
\begin{gathered}
\frac{3qg}{\sqrt{g^2+12q^2e^2}}{\rm
diag.}(0,0,1)
\hspace{0.2cm}\hbox{for polar and A}\\
\end{gathered}
\hfill\\
\begin{gathered}
\frac{3qg}{\sqrt{g^2+48q^2e^2}}{\rm
diag.}(1,1,-1)
\hspace{0.2cm}\hbox{for planar.}\\
\end{gathered}
\end{array}
\right.
\label{charge}
\end{equation}

The thermal equilibrium in a magnetic field $H\hat{\bf z}$ is
determined by minimizing the Gibbs free energy density,
\begin{equation}
{\cal G}=\Gamma-BH \label{gibbs}
\end{equation}
where $\Gamma$ is the thermodynamical potential in the grand canonical
ensemble. Ignoring the induced magnetization due to the normal current, we have
\begin{equation}
\Gamma = \frac{1}{2}B^2+\frac{1}{2}\sum_{l=1}^8(B^l)^2-p
\end{equation}
where $p$ is the pressure at $B=0$, maximized
with respect the gap parameter in the case of the CSC phase.
The minimization with respect to $B$ and $B^l$ in other CSC phases
is subject to the constraint imposed by the Meissner effect. For a hypothetical
quark matter of one flavor only, we find that
\begin{equation}
{\cal G}=\left\{\begin{array}{ll}
\begin{gathered}
-p_n-\frac{1}{2}H^2,
\hspace{0.2cm}\hbox{for normal phase}\\
\end{gathered}
\hfill\\
\begin{gathered}
-p_{\rm CSL},
\hspace{0.2cm}\hbox{for CSL}\\
\end{gathered}
\hfill\\
\begin{gathered}
-p_i-\frac{1}{2}H^2\cos^2\gamma_i,
\hspace{0.2cm}\hbox{for $i$=polar, A, planar}\\
\end{gathered}
\end{array}
\right.
\label{minimum}
\end{equation}
after the minimization. As will be shown below,
\begin{equation}
p_n<p_{\rm A}<p_{\rm polar}<p_{\rm planar}<p_{\rm CSL}.
\label{inequality}
\end{equation}
The phase corresponding to minimum among ${\cal G}$'s above wins the
competition and transition from one phase to another is first order
below $T_c$.

The situation becomes more subtle when quarks of different flavors
coexist even though pairing is within each flavor. Different
electric charges of different quark flavors imply different mixing
angles which may not be compactible with each other. Consider a
quark matter of u and d flavors with each flavor in a nonspherical
CSC phase with different mixing angles. Eq.(\ref{constraint})
imposes two constraints, which are consistent with each other only
if $B=B^8=0$.Then we end with an effective Meissner
shielding\cite{SWR}, making it fail to compete with the phase of
both flavors in CSL states.  On the other hand, one may relax the
constraints by assuming that the basis underlying the CSC phase of u
quarks differ from that underlying the CSC of d quarks by a color
rotation. Consequently the constraint (\ref{constraint}) for each
flavor yields $B^8=-B\tan\gamma^u$ and $B^{\prime8}=-B\tan\gamma^d$.
If both flavors stay in the polar or planar  phases, which allows
${\bf B}^{1-3}$ to penetrate in, an orthogonal transformation
\begin{eqnarray}
B^{\prime 8}&=&B^8\cos\beta-B^3\sin\beta\nonumber\\
B^{\prime 3}&=&B^8\sin\beta+B^3\cos\beta
\end{eqnarray}
could compromise both constraints. Such a transformation, however,
cannot be implemented in an adjoint representation of the color
$SU(3)$ and therefore, the mutual rotation of color basis is not an
option. The phases of  the two flavor quark matter (u,d) without
Meissner effects, which can compete with  (CSL,CSL), include
(polar,planar), (polar(normal), normal(polar)), (A(normal),
normal(A)) and (normal, normal). Notice the coincidence of the
mixing angle of the polar phase of u quarks and that of the planar
phase of the d quarks. Also the normal phase does not impose any
constraint on the gauge field and can coexist with any nonspherical
CSC.

The Gibbs free energies of (normal, normal) and (CSL, CSL) phases
remain given by the first and the second equations of
(\ref{minimum}), but with $p_n$ and $p_{\rm CSL}$ referring to the
total pressure of u and d quarks. For nonspherical phases, we have
\begin{equation}
{\cal G}=-p-\frac{1}{2}H^2\cos\gamma.
\end{equation}
where $p$ is the total pressure of both flavors with at least one
of them in a nonspherical CSC state and $\gamma$ is their common
mixing angle. For normal-CSC combination, $\gamma$ refers to that
of the CSC state. The charge neutrality condition is imposed in
all phases, which makes the Fermi sea of d quarks larger than that
of u quarks. The color neutrality condition is ignored owing to
small energy gap associated to the single flavor pairing. The number of
combinations to be examined is reduced by two criteria: 1) For two
combinations of the same mixing angle, the one with higher
pressure wins. 2) For two combinations of the same pressure, the
one with smaller magnitude of the mixing angle wins. It follows
that there are only four phases to be considered in each case of
two and three flavors, which are shown in Table I. The phase diagram
of each case in ultrarelativistic limit will be determined below and
their relevance to the realistic s quark mass will be discussed
afterwards.


The pressure of the single flavor CSC in the absence of a magnetic
field has been obtained in the literature at zero temperature
within the frame work of the one-gluon-exchange. We shall extend
the analysis up to the transition temperature
$T_c$, which is universal for all single flavor pairings. To avoid
the technical complexity of the one-gluon-exchange, we shall work
with a NJL-like effective action which picks up only the dominant
pairing channel of the former, the transverse pairing, in the
ultra-relativistic limit. The Hamiltonian of the effective action
reads\cite{Alford}
\begin{equation}
{\cal H}=\int d^3{\bf
r}\Big[\bar\psi(\vec\gamma\cdot\vec\nabla-\mu\gamma_4)\psi
-G\bar\psi\gamma_\mu T^l\psi\bar\psi\gamma_\mu T^l\psi \Big].
\end{equation}
with $T^l=\frac{1}{2}\lambda^l$ and $G$ an effective coupling.
Introducing the condensate (\ref{diquark}), we find the pressure of
each flavor under mean field approximation
\begin{eqnarray}
\nonumber p=&-&\frac{2}{\Omega}\sum_{{\bf k}}(k-\mu-E_{\bf k})
-\frac{1}{\Omega}\sum_{\bf
k}(k-\mu-|k-\mu|)\\
\nonumber &+&\frac{2T}{\Omega}\sum_{\bf
k}\ln\left(1+e^{-\frac{|p-\mu|}{T}}\right)-\frac{9}{4G}\Delta^2\\
&+&\frac{4T}{\Omega} \sum_{\bf k}\ln\left(1+e^{-\frac{E_{\bf
k}}{T}}\right), \label{pressure}
\end{eqnarray}
where $E_{\bf k}=\sqrt{(k-\mu)^2+\Delta^2f^2(\theta)}$ with
$\theta$ the angle between ${\bf k}$ and a prefixed spatial
direction and $\Delta$ given by the solution of the gap equation
$\left(\frac{\partial p}{\partial\Delta}\right)_\mu=0$. The
function $f(\theta)$ is given by
\begin{equation}
f(\theta)=\left\{\begin{array}{ll}
\begin{gathered}
1,
\hspace{0.2cm}\hbox{for CSL phase}\\
\end{gathered}
\hfill\\
\begin{gathered}
\sqrt{\frac{3}{4}(1+\cos^2\theta)},
\hspace{0.2cm}\hbox{for planar phase}\\
\end{gathered}
\hfill\\
\begin{gathered}
\sqrt{\frac{3}{2}}\sin\theta,
\hspace{0.2cm}\hbox{for polar phase}\\
\end{gathered}
\hfill\\
\begin{gathered}
\sqrt{3}\cos^2\frac{\theta}{2}.
\hspace{0.2cm}\hbox{for A phase}\\
\end{gathered}
\end{array}
\right.
\label{function}
\end{equation}
Introducing $\Delta p_s\equiv
p_s-p_n\equiv\rho_s(T)\frac{\mu^2\Delta_0^2}{2\pi^2}$ with $s$
labeling different pairing states and $\Delta_0$ the CSL gap at
$T=0$. We have $\rho_{\rm CSL}(0)=1$, $\rho_{\rm planar}(0)=0.98$,
$\rho_{\rm polar}(0)=0.88$ and $\rho_{\rm A}(0)=0.65$, and
$\rho_s(T_c)=0$ with $T_c=\frac{e^{\gamma_E}}{\pi}\Delta_0$. The
function $\rho_s(T)$ for $0<T<T_c$ of various states are displayed
in Fig.1, which satisfy the inequalitis (\ref{inequality}).

\begin{figure}
\includegraphics[height=2.5in, width=3.5in]{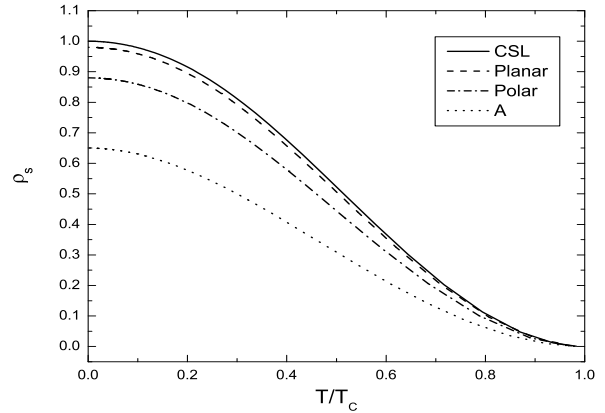}
\caption{\label{fig:epsart} The function $\rho_s(T)$ for various
pairing states.}
\end{figure}

In a multiflavor quark matter, the Fermi momentum of each flavor is
displayed from each other to meet the charge neutrality requirement.
For an ideal gas of (u,d) quarks and electrons at zero temperature,
we find that $k_u= 0.87\mu$ and $k_d=1.09\mu$. While for an ideal
gas of (u,d,s) quarks and electrons with $m_s<<\mu$, we obtain that
$k_u=\mu$, $k_d=\mu+\frac{m_s^2}{4\mu}$ and
$k_s=\mu-\frac{m_s^2}{4\mu}$.  The corrections brought about by
nonzero temperature and/or gap parameters contribute a higher order
term than $O(\mu^2\Delta^2)$ to the pressure and can be neglected
here.


By balancing the Gibbs free energy of different phases, we obtain
the phase diagram with respect to temperature and magnetic field.
The two flavor and three flavor cases are shown in Fig.2, where
$H_0$ is defined by
\begin{equation}
H_0=\frac{\mu\Delta_0}{\pi}.
\end{equation}
If we calibrate the effective coupling $G$ by identifying
$\Delta_{0}$ with that of the one-gluon exchange\cite{T, gap}
\begin{equation}
\Delta_0=512\pi^4\left(\frac{2}{N_f}\right)^{\frac{5}{2}}
\frac{\mu}{g^5}\exp\left(-\frac{3\pi^2}{\sqrt{2}g}-\frac{\pi^2+4}{8}-\frac{9}{2}\right)
\end{equation}
extrapolated to $\mu=500$MeV and $\alpha_s=1$, we end up with the values
of $H_0$ and $T_c$ in Table. I. For the three flavor case, we ignored
the Fermi momentum mismatch to be consistent with the ultra-relativistic
approximation.

A critical reader may challenge our ultra-relativistic treatment of
s quarks in the three flavor case, which may be justified as follows: The maximum Fermi-momentum mismatch supporting a cross-flavor pairing
scales with the energy gap, which is much smaller than the chemical
potential. The realistic value of the s qurak mass ($\simeq$150MeV),
could induced a sustantial mismatch that suppresses cross flavor
pairings, but remains small enough to warrant an ultra-relativistic
approximation of the pairing dynamics. Therefore we argue that
the three flavor panel of Fig.2 captured the gross features of the
phase diagram with the realistic s quark mass. We also admit that
the approximation may be marginal and perturbation of the mass
will be considered in near future.


\begin{figure}
\includegraphics[height=4.0in, width=3.6in]{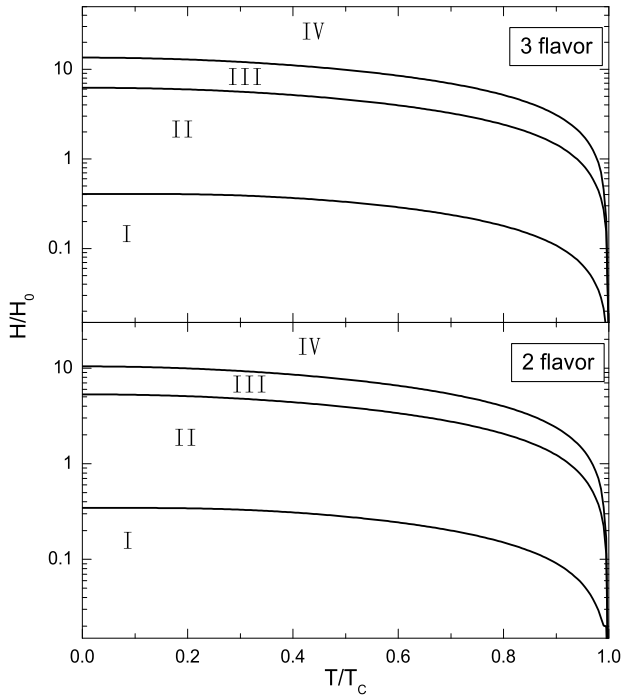}
\caption{\label{fig:epsart} H-T phase diagram for two flavors and
three flavors.}
\end{figure}

The analysis up to now ignores the magnetization $M=\frac{\partial
p}{\partial B}$\cite{fn2} in the absence of the Meissner effect and
we attempt to justify this approximation here. The potential hazard
comes from the de Hass-van Alphen (dHvA) effect stemming from the
discreteness of the Landau orbits if the mean free path $l$ of
quarks is longer than the cyclotron radius, $\mu/(eB)$. Even though
the magnetic field in the phase diagrams is weak in the sense
$(eB)^2<<\mu$, a large magnetization may emerge through the
derivative because of the rapid oscillation. The stability condition
$\frac{\partial^2{\cal G}}{\partial B^2}>0$, however, prevents its
happening. Along the equilibrium M-B curve constructed by the
Maxwell rule, the ratio $M/B$ cannot exceed the order of
$\alpha_e^{2/3}$ in the normal phase. This is also expected to be
the case in a nonspherical CSC phase. Because of the nonzero charges
of the pairing partners Eq.(\ref{charge}), the Laudau orbits also
impacts on the energy gap in the planar phase and a similar issue
for CFL has been addressed numerically in the
literature\cite{FIM,KW,NS}. Our analytic work reveals that the
magnitude of the oscillatory term of the gap is suppressed by
$O(\sqrt{eB}/\mu)$ relative to the term at $B=0$. In the opposite
limit where $l<<\mu/(eB)$, the dHvA oscillation is smeared out by
scattering.

To conclude, we have explored the consequences of the absence of the
electromagnetic Meissner effect in a nonspherical CSC phase of
single flavor pairing. We found that these nonspherical phases
occupy a significant portion of the $H-T$ phase diagram for the
plausible magnitude of the magnetic field inside a compact star. The
physical implications of these possible phases  to  the  cooling
behaviors and r-mode instability of neutron stars are interesting
topics deserving further investigations\cite{Alford:2006vz} .

The nonspherical phases discussed in this paper are all homogeneous
in space. A domain wall structure was suggested in \cite{Son} in the
context of 2SC and CFL in a magnetic field. The mechanism involves
the absence of the Meissner effect, the chiral symmetry breaking and
the axial anomaly. Since the transverse pairing, which pairs quarks
of opposite helicities, also breaks the chiral symmetry, it would be
interesting to extend the analysis of \cite{Son} to the nonspherical
phases.  We have not considered  the noninert phases discussed
in\cite{B}. But in any case, the importance of the nonspherical CSC
in a magnetic field, revealed in this letter, will stand up.

\begin{acknowledgments}
We would like to extend our gratitude to  D. Rischke, T.
Sch$\ddot{a}$fer,A. Schmitt, I. Shovkovy, V. Incera, E. Ferrer, Q.
Wang and X.P. Zheng for helpful discussions. We especially thank A.
Schmitt for pointing out a numerical error. The work of D. F. H. and
H. C. R. is supported in part by NSFC under grant Nos. 10575043,
10735040.

\end{acknowledgments}

\end{document}